\documentclass{nature}

\usepackage{graphicx}
\usepackage{amsmath}
\usepackage{amssymb}
\usepackage{color}
\usepackage{dcolumn}
\usepackage{epsfig}
\usepackage{bm}
\usepackage{multirow}
\usepackage{supertabular}
\usepackage[urlcolor=blue]{hyperref}
\hypersetup{backref, colorlinks=true, linkcolor=blue, citecolor=blue}

\title{Superconductivity in an organometallic compound.}

\author{Ren-Shu Wang$^{1,2}$, Liu-Cheng Chen$^1$, Hui Yang$^2$, Ming-An Fu$^2$, Jia Cheng$^2$, Xiao-Lin Wu$^2$, Yun Gao$^2$, Zhong-Bing Huang$^2$ \& Xiao-Jia Chen$^1$}

\begin{document}

\maketitle

\begin{affiliations}
 \item Center for High Pressure Science and Technology Advanced Research, Shanghai 201203, China
 \item Faculty of Materials Science and Engineering, Faculty of Physics and Electronic Technology, Hubei University, Wuhan 430062, China
\end{affiliations}

\begin{abstract}

Organometallic compounds constitute a very large group of substances that contain at least one metal-to-carbon bond in which the carbon is part of an organic group. They have played a major role in the development of the science of chemistry. These compounds are used to a large extent as catalysts (substances that increase the rate of reactions without themselves being consumed) and as intermediates in the laboratory and in industry. Recently, novel quantum phenormena such as topological insulators and superconductors were also suggested in these materials. However, there has been no report on the experimental exploration for the topological state. Evidence for superconductivity from the zero-resistivity state in any organometallic compound has not been achieved yet, though much efforts have been devoted. Here we report the experimental realization of superconductivity with the critical temperature of 3.6 K in a potassium-doped organometallic compound, {\it i.e.} tri-\emph{o}-tolylbismuthine with the evidence of both the Meissner effect and the zero-resistivity state through the $dc$ and $ac$ magnetic susceptibility and resistivity measurements. The obtained superconducting parameters classify this compound as a type-II superconductor. The benzene ring is identified to be the essential superconducting unit in such a phenyl organometallic compound. The superconducting phase and its composition are determined by the combined studies of the X-ray diffraction and theoretical calculations as well as the Raman spectroscopy measurements. These findings enrich the applications of organometallic compounds in superconductivity and add a new electron-acceptor family for organic superconductors. This work also points to a large pool for finding superconductors from organometallic compounds.
\end{abstract}

The organometallic compounds first discovered by Zeise of Denmark in 1827, have been an active frontier in organic, inorganic, coordination, and biological chemistry\cite{1}. These compounds usually contain at least one chemical bond connecting a carbon (C) atom of an organic molecule and a metal element (M). Their extensive prospects have been found in the applications for catalyzer\cite{2,21} (most famously, Ziegler$-$Natta catalyst), organic synthesis reagents\cite{3,31} (most commonly, Grignard reagent), anticancer drugs ($e.g.$ metallocene dihalides\cite{4}), and other special reagents\cite{51} (such as antiseptic, sterilizer, and insecticide). However, the applications of their physical properties have not widely been addressed as expected. Recently, these compounds were suggested theoretically as candidates of topological insulators\cite{Wang2015Organic} due to their lattice symmetry and strong spin-orbit coupling. In general, many topological insulators can become superconductors by chemical doping\cite{hor} or the application of external pressure\cite{kirs}. It is interesting to examine whether such organometallic compounds could eventually exhibit superconducting properties through some external modifications.  

As a typical organometallic compound, tri-\emph{o}-tolylbismuthine (\emph{o}-TTB) with the structure of three methyl connected to triphenylbismuth (TPB) in the ortho-position of the C-Bi bond is one of the inexpensive and nontoxic organobismuth reagents. These reagents are widely used in the chemical synthesis such as in the preparation of transition metal complexes or as catalysts for polymerization reaction in medicinal chemistry\cite{7}. Very recently, Meissner effect was reported in potassium (K)-doped TPB\cite{renshu}, though the evidence for superconductivity from electrical transport measurements is still absent. Differing with TPB, one surrounding hydrogen in each phenyl is replaced by methylphenyl in \emph{o}-TTB. Although both compounds still share similar physicochemical properties, adding CH$_3$ in \emph{o}-TTB affects the molecular geometry due to the different steric hindrance. This in turn induces additional C-H intermolecular interactions. Previous studies\cite{Soyoun} have shown that the Bi-Bi distance increases from 5.11 to 5.71 {\AA} from TPB to \emph{o}-TTB. This may facilitate metal-doping in \emph{o}-TTB molecules from spatial arrangement perspective. On the other hand, the addition of methyl generally reduces the symmetry because of the introduction of the lattice distortion. As a result, the symmetry space group changes from \emph{C}2/c for monoclinic TPB to \emph{P}1 for triclinic \emph{o}-TTB. This degradation in structural symmetry as well as intercalation of metal atoms can lead to the increase of carrier concentration which helps to enhance the electrical conductivity. Therefore, \emph{o}-TTB is expected to exhibit better electrical transport properties by a simple addition of the functional groups. Here we choose an \emph{o}-TTB molecule to examine possible superconductivity by doping alkali metal.

Meissner effect and zero-resistance are two essential characters for superconductivity. The Meissner effect in K-doped \emph{o}-TTB is comfirmed by both the $dc$ and $ac$ magnetic susceptibility ($\chi$) measurements. Figure 1a shows the temperature dependence of the $dc$ magnetic susceptibility ($\chi_{dc}$) with the zero-field cooling (ZFC) and field cooling (FC) runs at the magnetic field of 10 Oe. One can see a sudden drop in $\chi_{dc}$ around 3.6 K in both the ZFC and FC runs. This sudden drop of $\chi_{dc}$, originating from the well-defined Meissner effect, manifests the occurrence of superconductivity in the studied organometallic compound. The $T_c$ is defined as the temperature where the sharp drop takes place. The shielding fraction extracted from the $\chi_{dc}$ at temperature of 1.8 K is about 17\%. This is around five times higher than the reported one (3.74\%) in K-doped TPB\cite{renshu}. This significant improvement could benefit from the contributions from the functional group $-$CH$_3$. Figure 1c shows the magnetic hysteresis (\emph{M-H}) loop with the scanning magnetic field up to 2 kOe at the temperature range of 2-3 K. The clear diamond-like hysteresis loop shrinks with increasing temperature. This behavior coincides with the feature of a typical type-II superconductor. The \emph{M} decreases linearly with increasing \emph{H} in the initial stage until the applied field \emph{H} exceeding the lower critical field ($H_{c1}$). The temperature-dependent $H_{c1}$ is shown in Fig. 1d. The inset of Fig. 1d shows the method for determining the value of $H_{c1}$ at 2 K, namely the field deviated from the linear behavior. The $H_{c1}(0)$ value of 62 $\pm$ 2 Oe is thus obtained from the extrapolation to zero temperature based on the empirical law $H_{c1}(T)/H_{c1}(0) = 1-(T/T_c)^2$. In Fig. 1b, the diamagnetic signal in the Meissner state is gradually suppressed until completely being faded by the applied magnetic field up to 2 kOe. This is an intrinsic property of a type-II superconductor.

The $ac$ magnetic susceptibility $\chi_{ac}$ measurements provide the further evidence for the occurrence of the Meissner effect, as shown in Fig. 1e. As a common technology for the identification of the existence of superconductivity, the real component $\chi_{ac}^{\prime}$ of the $ac$ susceptibility reveals the magnetic shielding, and the imaginary part $\chi_{ac}^{\prime\prime}$ is a measure of the magnetic irreversibility\cite{Hein}. Both components of $\chi_{ac}$ exhibit exquisite changes at around 3.6 K. This temperature is exactly the same as that detected from the temperature dependence of $\chi_{dc}$. Since the absence of the magnetic flux in the normal state, the values of both parts are close to zero. Below $T_c$, the diamagnetic signal in $\chi_{ac}^{\prime}$ drops fast with lowering temperature due to the exclusion of magnetic flux. In imaginary part, the flux penetrating sample falls behind the applied flux to form a positive peak signal. Such a peak signal in $\chi_{ac}^{\prime\prime}$ implies the tendency to form the zero-resistivity in the superconducting state at a qualitative level.

Now we check the existence of the zero-resistivity in K-doped \emph{o}-TTB. Figure 2a displays the resistivity measurements on this compound at ambient pressure. It can be seen that the change of the resistivity with temperature is not regular. There seemingly exists a hump at around 120 K. Above that, the resistivity shows non-metallic feature. It changes to metallic-like behavior at low temperatures. This phenomenon is probably due to the weak linkage inside the sample or the poor contact between the sample and electrode. However, as temperature is decreased, the resistivity suddenly shows a sharp drop and then gets to zero at a certain temperature. This observation serves the solid evidence for supporting the existence of superconductivity in this organometallic compound. The onset temperature, below which a sudden drop of the resistivity is observed, is just the critical temperature for this new superconductor. Its value is exactly the same as that detected from the magnetic susceptibility measurements (Fig. 1). The inset of Fig. 2a shows the enlarged view of the low temperature resistivity. The zero-resistivity behavior with $T_c$ $\sim$ 3.6 K can be clearly observed. Therefore, K-doped \emph{o}-TTB is identified as a new organic superconductor from both the Meissner effect and zero-resistivity state. In comparison with TPB\cite{renshu}, adding $-$CH$_3$ group in \emph{o}-TTB does not make much difference for $T_c$ rather than yielding a huge uptake for superconducting shield fraction. This indicates that the phenyl is the essential unit for holding superconductivity in phenyl organometallic compounds. The enhancement of the superconducting shield fraction is a key factor for the realization of the zero-resistivity state in K-doped \emph{o}-TTB. Therefore, inducing the functional groups offers a simple but effective means to improve superconducting properties in organometallic compounds.

The superconducting parameters can be drawn from the field dependences of both the magnetic susceptibility and resistivity. Figure 2b shows the temperature dependence of the resistivity at pressure of 5 kbar and at various magnetic fields up to 1.3 Tesla. The suppression of superconductivity can be found by the application of magnetic fields. The temperature-dependent resistivity curve systematically shifts toward lower temperatures with increasing magnetic fields. Meanwhile, the temperature span of superconducting transition broadens significantly as the magnetic field is increased. Superconductivity is completely destroyed at the magnetic field of 1.3 Tesla in the studied temperature range. The temperature-dependent resistivity curves at various magnetic fields allow the determination of an important superconducting parameter $-$ the upper critical field ($H_{c2}$). $H_{c2}$ is defined by using the onset $T_c$ criteria, which is determined by the first dropped point deviated from the linear resistivity curve. The inset of Fig. 2b summarizes the temperature dependence of $H_{c2}$. Based on the Werthamer-Helfand-Hohenberg equation\cite{BCS}: $H_{c2}(0) = 0.693[-(dH_{c2}/dT)]_{T_c}T_c$, one can obtain the value of $H_{c2}(0)$ for the compound at pressure of 5 kbar. The calculated $H_{c2}(0)$ is about 1.78 $\pm$ 0.10 Tesla at 0 K. The colorful area is extrapolated by using the formula of $H_{c2}(T) = H_{c2}(0)[1-(T/T_c)^2]/[1+(T/T_c)^2]$. Since the $T_c$ values at the ambient pressure and at pressure of 5 kbar are almost the same, we can assume the same $H_{c2}(0)$ value for the ambient condition. By using the equations for the critical fields\cite{M} $H_{c2}(0) =\Phi_{0}/2\pi\xi_{0}^2$ and $H_{c1}(0) = (\Phi_{0}/4\pi\lambda_{L}^2)\ln(\lambda_{L}/\xi_{0})$ with $\Phi_{0}$ being the flux quantum, one obtains the zero-temperature superconducting coherence length $\xi_{0}$ of 136 $\pm$ 3 {\AA} and the London penetration depth $\lambda_{L}$ of 2840 $\pm$ 4 {\AA}. The Ginzburg-Landau parameter $\kappa$ =  20.9 $\pm$ 0.4 is thus obtained based on the expression $\kappa = \lambda_{L}/\xi_{0}$, supporting the feature of a type-II superconductor\cite{M}. The obtained superconducting parameters in K-doped \emph{o}-TTB are reasonably comparable to those for the low-dimensional organic salts\cite{L} and metal-doped fullerides\cite{P}.

One may wonder what the superconducting phase of K-doped \emph{o}-TTB could be. The crystal structures of pristine and K-doped \emph{o}-TTB are showed in Fig. 3. The powder X-ray diffraction (XRD) patterns of pure \emph{o}-TTB (Fig. 3a) can be indexed well as a triclinic class with the space group of \emph{P}1(1). There are eight molecules of C$_{21}$H$_{21}$Bi in a unit cell with the lattice parameters  $a$ = 38.3100 {\AA}, $b$ = 5.2500 {\AA}, and $c$ = 20.2200 {\AA} together with angles $\alpha$ = 90.00$^\circ$, $\beta$ = 121.00$^\circ$, and $\gamma$ = 90.00$^\circ$, as shown in Fig. 3c. The crystal structure of this organometallic compound changes dramatically after doping alkali metal (Fig. 3b), implying the formation of a distinct structure after potassium joining in. The crystal structure of K-doped \emph{o}-TTB is obtained on the basis of K-doped TPB\cite{renshu}. Both compounds show very similar XRD patterns (Fig. 3b). While replacing TPB with \emph{o}-TTB in the unit cell of K-doped TPB, we performed the full relaxation of the atomic positions for the mole ratio of K and \emph{o}-TTB in 4:1 and 3:1. The optimized results showed that the XRD patterns in the ratio of 3:1 fairly match with the measurements. For such a structure, three molecules of C$_{21}$H$_{21}$Bi and nine K atoms distribute in a nearly cubic unit cell with the dimensions $a$ = 9.5450 {\AA}, $b$ = 9.5810 {\AA}, and $c$ = 9.5530 {\AA} together with angles $\alpha$ = 89.62$^\circ$, $\beta$ = 90.39$^\circ$, and $\gamma$ = 89.85$^\circ$, as shown in Fig. 3d. Here K atoms represented by blue balls are intercalated in the interstitial space of bismuth and methylphenyl rings. In addition, the metal Bi trace is also observed in XRD patterns of doped sample (marked by the symbol $\ast$), indicating partial decomposition of \emph{o}-TTB into Bi atoms and methylbenzene (colorless liquid, removed). Similar situation has been discussed in K-doped TPB\cite{renshu}. Therefore, the K$_3$\emph{o}-TTB is the most possible superconducting phase from the comparison of the experimental observation and theoretical calculations. The atomic positions of C and Bi in \emph{o}-TTB and those of C, H, Bi, and K in K$_{3}$\emph{o}-TTB are given in the Supplementary Information.

Raman spectroscopy measurements were performed to understand the formation of the superconducting phase. Figure 4 shows Raman spectra of pristine and doped \emph{o}-TTB in the frequency range of 0-1800 cm$^-$$^1$. Pure \emph{o}-TTB (red curve) displays four regions of the vibrational modes for the lattice and Bi-phenyl, C-C-C bending, C-H bending, and C-C stretching, from low to high frequencies\cite{Ludwig1995Fourier}. By intercalating K into \emph{o}-TTB, the lattice and C-C-C bending regions vanish abruptly. In addition, the other zones (C-H bending and C-C stretching) change dramatically. In C-H bending region, three Raman peaks (637, 1010 and 1197 cm$^-$$^1$) for \emph{o}-TTB exhibit distinct red shift upon dopant. This indicates a phonon-mode softening effect, arising from charge-transfer between alkali metal and organobismuth molecule. This downshift effect has been generally observed in C-bearing superconductors such as alkali-metal doped hydrocarbons\cite{Mitsuhashi2010Superconductivity,Wang2012Superconductivity,Qiao2014Constraint}. This has been widely accepted as an approach to determine actual doping concentration in these superconductors. For K-doped \emph{o}-TTB, there are 21-23 cm$^-$$^1$ redshift in frequencies for the mentioned three phonon modes. The downshift with 6 or 7 cm$^-$$^1$ in the Raman spectra usually corresponds to one electron transfer. The redshift in our K-doped \emph{o}-TTB gives the amount of the transferred electrons of about 3. This is in good agreement with the result determined from the analysis of the XRD data (Fig. 3). On the other hand, the frequencies in the C-C stretching region increase with K doping. This behavior results from the phenyl polarization when benzene ring is connected to metal\cite{Postmus1969Vibrational}. Therefore, the complicated behavior of the Raman spectra of this superconductor is the product of the competition between the charge-transfer effect and benzene polarization.

The discovery of superconductivity in K-doped \emph{o}-TTB enriches the physical properties and adds the potential technological applications of organometallic compounds as superconductors. The introduction of methyl in pristine organometallic molecules leads to a nearly five-fold increase of superconducting shield fraction compared to K-doped TPB\cite{renshu} due to the possible delocalization effect and/or the increase of the carrier concentration. This offers an effective method to tune (super)conductivity by the simple addition of the functional groups. Our findings add a new electron-acceptor family for organic superconductors distinguished from M(dmit)$_2$ (M= Ni or Pd) system\cite{First}, polyaromatic hydrocarbons\cite{Mitsuhashi2010Superconductivity,Wang2012Superconductivity}, and \emph{p}-oligophenyls\cite{wang1,wang2,wang3,G.Zhong,jiafeng,gehuang}. This work also points to a new pool for producing superconductors from organometallic compounds.

\newpage

\noindent\textbf{Supplementary Note 1: The atomic positions of C and Bi in \emph{o}-TTB:}

\begin{center}
\topcaption{\label{tab:table1}The atomic positions of C and Bi in \emph{o}-TTB.}
\setlength{\tabcolsep}{7mm}
\begin{supertabular}{c c c c c c}
\hline
atom & Wyck. & \emph{x}/\emph{a}     & \emph{y}/\emph{b}     & \emph{z}/\emph{c}     & \emph{U} ({\AA}$^2$) \\ \hline \hline
C1   & 1a    & 0.32481 & 0.22148 & 0.9646  & 0.0127 \\
C2	 & 1a	 & 0.67519 & 0.77852 & 0.0354  & 0.0127 \\
C3	 & 1a	 & 0.67519 & 0.22148 & 0.5354  & 0.0127 \\
C4	 & 1a	 & 0.32481 & 0.77852 & 0.4646  & 0.0127 \\
C5	 & 1a	 & 0.82481 & 0.72148 & 0.9646  & 0.0127 \\
C6	 & 1a	 & 0.17519 & 0.27852 & 0.0354  & 0.0127 \\
C7	 & 1a	 & 0.17519 & 0.72148 & 0.5354  & 0.0127 \\
C8	 & 1a	 & 0.82481 & 0.27852 & 0.4646  & 0.0127 \\
C9	 & 1a	 & 0.47611 & 0.88163 & 0.14048 & 0.0127 \\
C10	 & 1a	 & 0.52389 & 0.11837 & 0.85952 & 0.0127 \\
C11	 & 1a	 & 0.52389 & 0.88163 & 0.35951 & 0.0127 \\
C12	 & 1a	 & 0.47611 & 0.11837 & 0.64048 & 0.0127 \\
C13  & 1a	 & 0.97611 & 0.38163 & 0.14048 & 0.0127 \\
C14	 & 1a	 & 0.02389 & 0.61837 & 0.85952 & 0.0127 \\
C15	 & 1a	 & 0.02389 & 0.38163 & 0.35951 & 0.0127 \\
C16  & 1a	 & 0.97611 & 0.61837 & 0.64048 & 0.0127 \\
C17	 & 1a	 & 0.49575 & 0.95862 & 0.102   & 0.0127 \\
C18	 & 1a	 & 0.50425 & 0.04138 & 0.898   & 0.0127 \\
C19	 & 1a	 & 0.50425 & 0.95862 & 0.398   & 0.0127 \\
C20	 & 1a	 & 0.49575 & 0.04138 & 0.602   & 0.0127 \\
C21	 & 1a	 & 0.99575 & 0.45862 & 0.102   & 0.0127 \\
C22	 & 1a	 & 0.00425 & 0.54138 & 0.898   & 0.0127 \\
C23	 & 1a	 & 0.00425 & 0.45862 & 0.398   & 0.0127 \\
C24	 & 1a	 & 0.99575 & 0.54138 & 0.602   & 0.0127 \\
C25	 & 1a	 & 0.47959 & 0.15589 & 0.04809 & 0.0127 \\
C26	 & 1a	 & 0.52041 & 0.84411 & 0.95191 & 0.0127 \\
C27	 & 1a	 & 0.52041 & 0.15589 & 0.45191 & 0.0127 \\
C28  & 1a	 & 0.47959 & 0.84411 & 0.54809 & 0.0127 \\
C29	 & 1a	 & 0.97959 & 0.65589 & 0.04809 & 0.0127 \\
C30	 & 1a	 & 0.02041 & 0.34411 & 0.95191 & 0.0127 \\
C31	 & 1a	 & 0.02041 & 0.65589 & 0.45191 & 0.0127 \\
C32	 & 1a	 & 0.97959 & 0.34411 & 0.54809 & 0.0127 \\
C33	 & 1a	 & 0.44453 & 0.28982 & 0.03326 & 0.0127 \\
C34	 & 1a	 & 0.55547 & 0.71018 & 0.96674 & 0.0127 \\
C35  & 1a	 & 0.55547 & 0.28982 & 0.46674 & 0.0127 \\
C36	 & 1a	 & 0.44453 & 0.71018 & 0.53326 & 0.0127 \\
C37	 & 1a	 & 0.94453 & 0.78982 & 0.03326 & 0.0127 \\
C38	 & 1a	 & 0.05547 & 0.21018 & 0.96674 & 0.0127 \\
C39	 & 1a	 & 0.05547 & 0.78982 & 0.46674 & 0.0127 \\
C40	 & 1a	 & 0.94453 & 0.21018 & 0.53326 & 0.0127 \\
C41	 & 1a	 & 0.4267  & 0.48608 & 0.9698  & 0.0127 \\
C42	 & 1a	 & 0.5733  & 0.51392 & 0.0302  & 0.0127 \\
C43	 & 1a	 & 0.5733  & 0.48608 & 0.5302  & 0.0127 \\
C44	 & 1a	 & 0.4267  & 0.51392 & 0.4698  & 0.0127 \\
C45	 & 1a	 & 0.9267  & 0.98608 & 0.9698  & 0.0127 \\
C46	 & 1a	 & 0.0733  & 0.01392 & 0.0302  & 0.0127 \\
C47	 & 1a	 & 0.0733  & 0.98608 & 0.5302  & 0.0127 \\
C48	 & 1a	 & 0.9267  & 0.01392 & 0.4698  & 0.0127 \\
C49	 & 1a	 & 0.37112 & 0.20502 & 0.15653 & 0.0127 \\
C50	 & 1a	 & 0.62888 & 0.79498 & 0.84346 & 0.0127 \\
C51	 & 1a	 & 0.62888 & 0.20502 & 0.34346 & 0.0127 \\
C52	 & 1a	 & 0.37112 & 0.79498 & 0.65653 & 0.0127 \\
C53	 & 1a	 & 0.87112 & 0.70502 & 0.15653 & 0.0127 \\
C54	 & 1a	 & 0.12888 & 0.29498 & 0.84346 & 0.0127 \\
C55	 & 1a	 & 0.12888 & 0.70502 & 0.34346 & 0.0127 \\
C56	 & 1a	 & 0.87112 & 0.29498 & 0.65653 & 0.0127 \\
C57	 & 1a	 & 0.39707 & 0.25083 & 0.23658 & 0.0127 \\
C58	 & 1a	 & 0.60293 & 0.74917 & 0.76342 & 0.0127 \\
C59	 & 1a	 & 0.60293 & 0.25083 & 0.26342 & 0.0127 \\
C60	 & 1a	 & 0.39707 & 0.74917 & 0.73658 & 0.0127 \\
C61	 & 1a	 & 0.89707 & 0.75083 & 0.23658 & 0.0127 \\
C62	&	1a	&	0.10293	&	0.24917	&	0.76342	&	0.0127	\\
C63	&	1a	&	0.10293	&	0.75083	&	0.26342	&	0.0127	\\
C64	&	1a	&	0.89707	&	0.24917	&	0.73658	&	0.0127	\\
C65	&	1a	&	0.39319	&	0.08887	&	0.28814	&	0.0127	\\
C66	&	1a	&	0.60681	&	0.91113	&	0.71186	&	0.0127	\\
C67	&	1a	&	0.60681	&	0.08887	&	0.21186	&	0.0127	\\
C68	&	1a	&	0.39319	&	0.91113	&	0.78814	&	0.0127	\\
C69	&	1a	&	0.89319	&	0.58887	&	0.28814	&	0.0127	\\
C70	&	1a	&	0.10681	&	0.41113	&	0.71186	&	0.0127	\\
C71	&	1a	&	0.10681	&	0.58887	&	0.21186	&	0.0127	\\
C72	&	1a	&	0.89319	&	0.41113	&	0.78814	&	0.0127	\\
C73	&	1a	&	0.36426	&	0.89379	&	0.26193	&	0.0127	\\
C74	&	1a	&	0.63574	&	0.10621	&	0.73806	&	0.0127	\\
C75	&	1a	&	0.63574	&	0.89379	&	0.23806	&	0.0127	\\
C76	&	1a	&	0.36426	&	0.10621	&	0.76193	&	0.0127	\\
C77	&	1a	&	0.86426	&	0.39379	&	0.26193	&	0.0127	\\
C78	&	1a	&	0.13574	&	0.60621	&	0.73806	&	0.0127	\\
C79	&	1a	&	0.13574	&	0.39379	&	0.23806	&	0.0127	\\
C80	&	1a	&	0.86426	&	0.60621	&	0.76193	&	0.0127	\\
C81	&	1a	&	0.34005	&	0.84373	&	0.18318	&	0.0127	\\
C82	&	1a	&	0.65995	&	0.15627	&	0.81682	&	0.0127	\\
C83	&	1a	&	0.65995	&	0.84373	&	0.31682	&	0.0127	\\
C84	&	1a	&	0.34005	&	0.15627	&	0.68318	&	0.0127	\\
C85	&	1a	&	0.84005	&	0.34373	&	0.18318	&	0.0127	\\
C86	&	1a	&	0.15995	&	0.65627	&	0.81682	&	0.0127	\\
C87	&	1a	&	0.15995	&	0.34373	&	0.31682	&	0.0127	\\
C88	&	1a	&	0.84005	&	0.65627	&	0.68318	&	0.0127	\\
C89	&	1a	&	0.28359	&	0.29754	&	0.92975	&	0.0127	\\
C90	&	1a	&	0.71641	&	0.70246	&	0.07025	&	0.0127	\\
C91	&	1a	&	0.71641	&	0.29754	&	0.57025	&	0.0127	\\
C92	&	1a	&	0.28359	&	0.70246	&	0.42975	&	0.0127	\\
C93	&	1a	&	0.78359	&	0.79754	&	0.92975	&	0.0127	\\
C94	&	1a	&	0.21641	&	0.20246	&	0.07025	&	0.0127	\\
C95	&	1a	&	0.21641	&	0.79754	&	0.57025	&	0.0127	\\
C96	&	1a	&	0.78359	&	0.20246	&	0.42975	&	0.0127	\\
C97	&	1a	&	0.34408	&	0.00052	&	0.13132	&	0.0127	\\
C98	&	1a	&	0.65592	&	-0.00052	&	0.86868	&	0.0127	\\
C99	&	1a	&	0.65592	&	0.00052	&	0.36868	&	0.0127	\\
C100	&	1a	&	0.34408	&	-0.00052	&	0.63132	&	0.0127	\\
C101	&	1a	&	0.84408	&	0.50052	&	0.13132	&	0.0127	\\
C102	&	1a	&	0.15592	&	0.49948	&	0.86868	&	0.0127	\\
C103	&	1a	&	0.15592	&	0.50052	&	0.36868	&	0.0127	\\
C104	&	1a	&	0.84408	&	0.49948	&	0.63132	&	0.0127	\\
C105	&	1a	&	0.43018	&	0.44808	&	0.26849	&	0.0127	\\
C106	&	1a	&	0.56982	&	0.55192	&	0.73151	&	0.0127	\\
C107	&	1a	&	0.56982	&	0.44808	&	0.23151	&	0.0127	\\
C108	&	1a	&	0.43018	&	0.55192	&	0.76849	&	0.0127	\\
C109	&	1a	&	0.93018	&	0.94808	&	0.26849	&	0.0127	\\
C110	&	1a	&	0.06982	&	0.05192	&	0.73151	&	0.0127	\\
C111	&	1a	&	0.06982	&	0.94808	&	0.23151	&	0.0127	\\
C112	&	1a	&	0.93018	&	0.05192	&	0.76849	&	0.0127	\\
C113	&	1a	&	0.25442	&	0.16023	&	0.86452	&	0.0127	\\
C114	&	1a	&	0.74558	&	0.83977	&	0.13548	&	0.0127	\\
C115	&	1a	&	0.74558	&	0.16023	&	0.63548	&	0.0127	\\
C116	&	1a	&	0.25442	&	0.83977	&	0.36452	&	0.0127	\\
C117	&	1a	&	0.75442	&	0.66023	&	0.86452	&	0.0127	\\
C118	&	1a	&	0.24558	&	0.33977	&	0.13548	&	0.0127	\\
C119	&	1a	&	0.24558	&	0.66023	&	0.63548	&	0.0127	\\
C120	&	1a	&	0.75442	&	0.33977	&	0.36452	&	0.0127	\\
C121	&	1a	&	0.2656	&	0.95797	&	0.83446	&	0.0127	\\
C122	&	1a	&	0.7344	&	0.04203	&	0.16553	&	0.0127	\\
C123	&	1a	&	0.7344	&	0.95797	&	0.66553	&	0.0127	\\
C124	&	1a	&	0.2656	&	0.04203	&	0.33447	&	0.0127	\\
C125	&	1a	&	0.7656	&	0.45797	&	0.83446	&	0.0127	\\
C126	&	1a	&	0.2344	&	0.54203	&	0.16553	&	0.0127	\\
C127	&	1a	&	0.2344	&	0.45797	&	0.66553	&	0.0127	\\
C128	&	1a	&	0.7656	&	0.54203	&	0.33446	&	0.0127	\\
C129	&	1a	&	0.30627	&	0.88213	&	0.87022	&	0.0127	\\
C130	&	1a	&	0.69373	&	0.11787	&	0.12978	&	0.0127	\\
C131	&	1a	&	0.69373	&	0.88213	&	0.62978	&	0.0127	\\
C132	&	1a	&	0.30627	&	0.11787	&	0.37022	&	0.0127	\\
C133	&	1a	&	0.80627	&	0.38213	&	0.87022	&	0.0127	\\
C134	&	1a	&	0.19373	&	0.61787	&	0.12978	&	0.0127	\\
C135	&	1a	&	0.19373	&	0.38213	&	0.62978	&	0.0127	\\
C136	&	1a	&	0.80627	&	0.61787	&	0.37022	&	0.0127	\\
C137	&	1a	&	0.33565	&	0.0149	&	0.93554	&	0.0127	\\
C138	&	1a	&	0.66435	&	-0.01491	&	0.06446	&	0.0127	\\
C139	&	1a	&	0.66435	&	0.0149	&	0.56446	&	0.0127	\\
C140	&	1a	&	0.33566	&	-0.01491	&	0.43554	&	0.0127	\\
C141	&	1a	&	0.83565	&	0.5149	&	0.93554	&	0.0127	\\
C142	&	1a	&	0.16434	&	0.4851	&	0.06446	&	0.0127	\\
C143	&	1a	&	0.16434	&	0.5149	&	0.56446	&	0.0127	\\
C144	&	1a	&	0.83565	&	0.4851	&	0.43554	&	0.0127	\\
C145	&	1a	&	0.27009	&	0.49959	&	0.9646	&	0.0127	\\
C146	&	1a	&	0.72991	&	0.50041	&	0.0354	&	0.0127	\\
C147	&	1a	&	0.72991	&	0.49959	&	0.5354	&	0.0127	\\
C148	&	1a	&	0.27009	&	0.50041	&	0.4646	&	0.0127	\\
C149	&	1a	&	0.77009	&	0.99959	&	0.9646	&	0.0127	\\
C150	&	1a	&	0.22991	&	0.00041	&	0.0354	&	0.0127	\\
C151	&	1a	&	0.22991	&	0.99959	&	0.5354	&	0.0127	\\
C152	&	1a	&	0.77009	&	0.00041	&	0.4646	&	0.0127	\\
C153	&	1a	&	0.42583	&	0.21763	&	0.07528	&	0.0127	\\
C154	&	1a	&	0.57417	&	0.78237	&	0.92472	&	0.0127	\\
C155	&	1a	&	0.57417	&	0.21763	&	0.42472	&	0.0127	\\
C156	&	1a	&	0.42583	&	0.78237	&	0.57528	&	0.0127	\\
C157	&	1a	&	0.92583	&	0.71763	&	0.07528	&	0.0127	\\
C158	&	1a	&	0.07417	&	0.28238	&	0.92472	&	0.0127	\\
C159	&	1a	&	0.07417	&	0.71763	&	0.42472	&	0.0127	\\
C160	&	1a	&	0.92583	&	0.28238	&	0.57528	&	0.0127	\\
C161	&	1a	&	0.44119	&	0.01362	&	0.12673	&	0.0127	\\
C162	&	1a	&	0.55881	&	-0.01362	&	0.87327	&	0.0127	\\
C163	&	1a	&	0.55881	&	0.01362	&	0.37327	&	0.0127	\\
C164	&	1a	&	0.44119	&	-0.01362	&	0.62673	&	0.0127	\\
C165	&	1a	&	0.9412	&	0.51362	&	0.12673	&	0.0127	\\
C166	&	1a	&	0.0588	&	0.48638	&	0.87327	&	0.0127	\\
C167	&	1a	&	0.0588	&	0.51362	&	0.37327	&	0.0127	\\
C168	&	1a	&	0.9412	&	0.48638	&	0.62673	&	0.0127	\\
Bi1	&	1a	&	0.37348	&	0.45428	&	0.06605	&	0.0127	\\
Bi2	&	1a	&	0.62652	&	0.54572	&	0.93395	&	0.0127	\\
Bi3	&	1a	&	0.62652	&	0.45428	&	0.43395	&	0.0127	\\
Bi4	&	1a	&	0.37348	&	0.54572	&	0.56605	&	0.0127	\\
Bi5	&	1a	&	0.87348	&	0.95428	&	0.06605	&	0.0127	\\
Bi6	&	1a	&	0.12652	&	0.04572	&	0.93395	&	0.0127	\\
Bi7	&	1a	&	0.12652	&	0.95428	&	0.43395	&	0.0127	\\
Bi8	&	1a	&	0.87348	&	0.04572	&	0.56605	&	0.0127	\\ \hline \hline
\end{supertabular}
\end{center}

\newpage

\noindent\textbf{Supplementary Note 2: The atomic positions of C, H, Bi, and K in K$_{3}$\emph{o}-TTB:}

\begin{center}
\topcaption {\label{tab:table2}The optimized atomic coordinates of C, H, Bi, and K for K$_{3}$\emph{o}-TTB.}
\setlength{\tabcolsep}{7mm}
\begin{supertabular}{c c c c c c}
\hline
atom & Wyck. & \emph{x}/\emph{a}     & \emph{y}/\emph{b}     & \emph{z}/\emph{c}     & \emph{U} ({\AA}$^2$) \\ \hline \hline
Bi1	&	1a	&	0.49625	&	0.49307	&	0.99311	&	0.0127	\\
Bi2	&	1a	&	0.49625	&	0.99161	&	0.50268	&	0.0127	\\
Bi3	&	1a	&	0	    &	0.49437	&	0.4925	&	0.0127	\\
K4	&	1a	&	0.00001	&	0.19529	&	0.20402	&	0.0127	\\
K5	&	1a	&	0.00001	&	0.79655	&	0.80298	&	0.0127	\\
K7	&	1a	&	0.197	&	0.19573	&	0.99165	&	0.0127	\\
K8	&	1a	&	0.197	&	0.99241	&	0.20165	&	0.0127	\\
K9	&	1a	&	0.80052	&	0.9908	&	0.80878	&	0.0127	\\
K10	&	1a	&	0.74814	&	0.24646	&	0.50177	&	0.0127	\\
K12	&	1a	&	0.5559	&	0.74378	&	0.26953	&	0.0127	\\
K13	&	1a	&	0.2278	&	0.74461	&	0.50491	&	0.0127	\\
K14	&	1a	&	0.74814	&	0.49489	&	0.25095	&	0.0127	\\
C16	&	1a	&	0.20306	&	0.37803	&	0.89211	&	0.0127	\\
C17	&	1a	&	0.3403	&	0.34292	&	0.94219	&	0.0127	\\
C18	&	1a	&	0.38378	&	0.20916	&	0.96676	&	0.0127	\\
C19	&	1a	&	0.29495	&	0.1037	&	0.94208	&	0.0127	\\
C20	&	1a	&	0.15267	&	0.12066	&	0.88642	&	0.0127	\\
C21	&	1a	&	0.11245	&	0.26251	&	0.86412	&	0.0127	\\
C22	&	1a	&	0.18385	&	0.7906	&	0.86277	&	0.0127	\\
C23	&	1a	&	0.18252	&	0.64137	&	0.86672	&	0.0127	\\
C24	&	1a	&	0.12802	&	0.5615	&	0.76202	&	0.0127	\\
C25	&	1a	&	0.07347	&	0.62448	&	0.64812	&	0.0127	\\
C26	&	1a	&	0.0706	&	0.77558	&	0.62496	&	0.0127	\\
C27	&	1a	&	0.12755	&	0.85372	&	0.73821	&	0.0127	\\
C28	&	1a	&	0.48936	&	0.15424	&	0.7637	&	0.0127	\\
C29	&	1a	&	0.55108	&	0.2423	&	0.86736	&	0.0127	\\
C30	&	1a	&	0.68148	&	0.29885	&	0.85523	&	0.0127	\\
C31	&	1a	&	0.75677	&	0.27258	&	0.7402	&	0.0127	\\
C32	&	1a	&	0.70783	&	0.1887	&	0.62219	&	0.0127	\\
C33	&	1a	&	0.57165	&	0.13119	&	0.64123	&	0.0127	\\
C34	&	1a	&	0.64006	&	0.65887	&	0.95866	&	0.0127	\\
C35	&	1a	&	0.75908	&	0.60382	&	0.88733	&	0.0127	\\
C36	&	1a	&	0.87	&	0.68337	&	0.84636	&	0.0127	\\
C37	&	1a	&	0.86973	&	0.82039	&	0.87465	&	0.0127	\\
C38	&	1a	&	0.75884	&	0.89268	&	0.95071	&	0.0127	\\
C39	&	1a	&	0.64511	&	0.80476	&	0.98966	&	0.0127	\\
C40	&	1a	&	0.09497	&	0.31777	&	0.3756	&	0.0127	\\
C41	&	1a	&	0.17219	&	0.18964	&	0.37569	&	0.0127	\\
C42	&	1a	&	0.26335	&	0.15358	&	0.2726	&	0.0127	\\
C43	&	1a	&	0.28384	&	0.24219	&	0.16535	&	0.0127	\\
C44	&	1a	&	0.21718	&	0.37933	&	0.15127	&	0.0127	\\
C45	&	1a	&	0.12206	&	0.41081	&	0.2611	&	0.0127	\\
C46	&	1a	&	0.65077	&	0.04323	&	0.32801	&	0.0127	\\
C47	&	1a	&	0.79981	&	0.0423	&	0.34464	&	0.0127	\\
C48	&	1a	&	0.89025	&	0.07037	&	0.23818	&	0.0127	\\
C49	&	1a	&	0.83851	&	0.10191	&	0.11086	&	0.0127	\\
C50	&	1a	&	0.68911	&	0.1121	&	0.07678	&	0.0127	\\
C51	&	1a	&	0.59966	&	0.08034	&	0.19148	&	0.0127	\\
C52	&	1a	&	0.67009	&	0.5544	&	0.55355	&	0.0127	\\
C53	&	1a	&	0.58751	&	0.50331	&	0.66715	&	0.0127	\\
C54	&	1a	&	0.48879	&	0.58169	&	0.73327	&	0.0127	\\
C55	&	1a	&	0.46573	&	0.71335	&	0.6901	&	0.0127	\\
C56	&	1a	&	0.53737	&	0.78045	&	0.57267	&	0.0127	\\
C57	&	1a	&	0.64022	&	0.69425	&	0.50852	&	0.0127	\\
C58	&	1a	&	0.83433	&	0.71856	&	0.3244	&	0.0127	\\
C59	&	1a	&	0.96444	&	0.64415	&	0.32549	&	0.0127	\\
C60	&	1a	&	0.0717	&	0.66184	&	0.2268	&	0.0127	\\
C61	&	1a	&	0.04882	&	0.75256	&	0.12142	&	0.0127	\\
C62	&	1a	&	0.91633	&	0.83871	&	0.10508	&	0.0127	\\
C63	&	1a	&	0.81521	&	0.81655	&	0.21205	&	0.0127	\\
C64	&	1a	&	0.3591	&	0.79776	&	0.32917	&	0.0127	\\
C65	&	1a	&	0.33175	&	0.70813	&	0.44608	&	0.0127	\\
C66	&	1a	&	0.34372	&	0.56634	&	0.44112	&	0.0127	\\
C67	&	1a	&	0.3843	&	0.50613	&	0.32161	&	0.0127	\\
C68	&	1a	&	0.42087	&	0.58231	&	0.19376	&	0.0127	\\
C69	&	1a	&	0.40509	&	0.72984	&	0.20454	&	0.0127	\\
H70	&	1a	&	0.17202	&	0.49213	&	0.87691	&	0.0127	\\
C71	&	1a	&	0.49435	&	0.18816	&	0.0147	&	0.0127	\\
H72	&	1a	&	0.32875	&	0.9845	&	0.9647	&	0.0127	\\
H73	&	1a	&	0.08534	&	0.02494	&	0.86504	&	0.0127	\\
H74	&	1a	&	0.00206	&	0.28175	&	0.82184	&	0.0127	\\
C75	&	1a	&	1.01798	&	0.81858	&	0.52157	&	0.0127	\\
C76	&	1a	&	0.38613	&	0.06689	&	0.76494	&	0.0127	\\
C77	&	1a	&	0.76454	&	0.48783	&	0.86196	&	0.0127	\\
H78	&	1a	&	0.96134	&	0.63331	&	0.78887	&	0.0127	\\
H79	&	1a	&	0.96032	&	0.88881	&	0.83868	&	0.0127	\\
H80	&	1a	&	0.76873	&	0.01594	&	0.97349	&	0.0127	\\
H81	&	1a	&	0.55521	&	0.85605	&	0.05528	&	0.0127	\\
H82	&	1a	&	0.35755	&	0.21239	&	0.07697	&	0.0127	\\
C83	&	1a	&	0.48206	&	0.08561	&	0.17037	&	0.0127	\\
C84	&	1a	&	0.38559	&	0.78107	&	0.74606	&	0.0127	\\
C85	&	1a	&	0.17452	&	0.60125	&	0.233	&	0.0127	\\
C86	&	1a	&	0.39213	&	0.38762	&	0.31497	&	0.0127	\\
H87	&	1a	&	0.84865	&	0.01718	&	0.45112	&	0.0127	\\
H88	&	1a	&	0.01563	&	0.06687	&	0.25626	&	0.0127	\\
H89	&	1a	&	0.9127	&	0.12294	&	0.02017	&	0.0127	\\
H90	&	1a	&	0.65416	&	0.14314	&	0.95954	&	0.0127	\\
H91	&	1a	&	0.56891	&	0.18053	&	0.98122	&	0.0127	\\
H92	&	1a	&	0.72518	&	0.36744	&	0.94217	&	0.0127	\\
H93	&	1a	&	0.86699	&	0.31681	&	0.72973	&	0.0127	\\
H94	&	1a	&	0.77729	&	0.17461	&	0.52669	&	0.0127	\\
H95	&	1a	&	0.22792	&	0.85142	&	0.95521	&	0.0127	\\
H96	&	1a	&	0.22862	&	0.58317	&	0.96084	&	0.0127	\\
H97	&	1a	&	0.12921	&	0.44287	&	0.77197	&	0.0127	\\
H98	&	1a	&	0.12683	&	0.97217	&	0.72609	&	0.0127	\\
C99	&	1a	&	0.15908	&	0.11151	&	0.46513	&	0.0127	\\
H100	&	1a	&	0.32097	&	0.0493	&	0.27825	&	0.0127	\\
H101	&	1a	&	0.24337	&	0.44973	&	0.05845	&	0.0127	\\
H102	&	1a	&	0.0663	&	0.51605	&	0.25468	&	0.0127	\\
H103	&	1a	&	0.34389	&	0.91532	&	0.33906	&	0.0127	\\
H104	&	1a	&	0.29804	&	0.75453	&	0.55017	&	0.0127	\\
H105	&	1a	&	0.31936	&	0.50056	&	0.53777	&	0.0127	\\
H106	&	1a	&	0.43115	&	0.79426	&	0.10745	&	0.0127	\\
H107	&	1a	&	0.75335	&	0.48472	&	0.50493	&	0.0127	\\
H108	&	1a	&	0.60257	&	0.39176	&	0.70683	&	0.0127	\\
H109	&	1a	&	0.40856	&	0.54972	&	0.78719	&	0.0127	\\
H110	&	1a	&	0.72001	&	0.77418	&	0.5052	&	0.0127	\\
H111	&	1a	&	0.75381	&	0.69797	&	0.40962	&	0.0127	\\
H112	&	1a	&	0.13101	&	0.76694	&	0.03655	&	0.0127	\\
H113	&	1a	&	0.89451	&	0.91747	&	0.0176	&	0.0127	\\
H114	&	1a	&	0.71517	&	0.88156	&	0.20507	&	0.0127	\\
H115	&	1a	&	0.573	&	0.2088	&	-0.07224	&	0	\\
H116	&	1a	&	0.50322	&	0.07475	&	0.05062	&	0	\\
H117	&	1a	&	0.5174	&	0.25992	&	0.10669	&	0	\\
H118	&	1a	&	0.98215	&	0.7279	&	0.45359	&	0	\\
H119	&	1a	&	1.10249	&	0.88009	&	0.46449	&	0	\\
H120	&	1a	&	0.92504	&	0.88994	&	0.54355	&	0	\\
H121	&	1a	&	0.71295	&	0.46862	&	0.75587	&	0	\\
H122	&	1a	&	0.71119	&	0.45884	&	0.94438	&	0	\\
H123	&	1a	&	0.8788	&	0.45316	&	0.85855	&	0	\\
H124	&	1a	&	0.4335	&	-0.02098	&	0.19156	&	0	\\
H125	&	1a	&	0.46254	&	0.11668	&	0.05648	&	0	\\
H126	&	1a	&	0.43253	&	0.16638	&	0.24195	&	0	\\
H127	&	1a	&	0.31302	&	0.71178	&	0.81028	&	0	\\
H128	&	1a	&	0.32137	&	0.8424	&	0.66545	&	0	\\
H129	&	1a	&	0.44222	&	0.85676	&	0.81842	&	0	\\
H130	&	1a	&	0.39722	&	0.32325	&	0.22804	&	0	\\
H131	&	1a	&	0.33303	&	0.33835	&	0.40599	&	0	\\
H132	&	1a	&	0.50691	&	0.35473	&	0.32042	&	0	\\
H133	&	1a	&	0.21068	&	0.00834	&	0.43608	&	0	\\
H134	&	1a	&	0.04278	&	0.09389	&	0.48484	&	0	\\
H135	&	1a	&	0.21066	&	0.15415	&	0.56391	&	0	\\
H136	&	1a	&	0.25042	&	0.64478	&	0.14591	&	0	\\
H137	&	1a	&	0.13644	&	0.53301	&	0.18773	&	0	\\
H138	&	1a	&	0.24892	&	0.65506	&	0.30918	&	0	\\
H139	&	1a	&	0.42361	&	-0.03752	&	0.80943	&	0	\\
H140	&	1a	&	0.31765	&	0.03673	&	0.66812	&	0	\\
H141	&	1a	&	0.29142	&	0.02718	&	0.82772	&	0	\\  \hline \hline
\end{supertabular}
\end{center}

\end{document}